\documentclass[prc,showpacs,showkeys,superscriptaddress,nofootinbib,twocolumn,floatfix]{revtex4-2}
\usepackage{graphicx,xcolor,amsmath,amssymb,amsthm,amsopn,bm,textcomp} 
\usepackage{cancel}

\newcommand{\Tr}{{\rm Tr}}
\newcommand{\vP}{\vec P}

\newcommand{\cG}{{\cal G}}

\newcommand{\ie}{\textit{i.e.}}
\newcommand{\eg}{\textit{e.g.}}

\newcommand{\QQb}{Q\bar{Q}}
\newcommand{\GQQb}{G_{Q\bar Q}}
\newcommand{\beq}{\begin{equation}}
\newcommand{\eeq}{\end{equation}}
\newcommand{\bea}{\begin{eqnarray}}
\newcommand{\eea}{\end{eqnarray}}

\usepackage{float}
\begin{document}

\title{Charmonia at Finite Momentum and Spatial Correlators in Quark-Gluon Plasma}

\author{T. Hardin}
\email[E-mail: ]{thomas$\_$hardin@tamu.edu}
\affiliation{Cyclotron Institute and Department of Physics and Astronomy, Texas A\&M University, College Station, Texas 77843-3366, USA} 

\author{R. Rapp} 
\email[E-mail: ]{rapp@comp.tamu.edu}
\affiliation{Cyclotron Institute and Department of Physics and Astronomy, Texas A\&M University, College Station, Texas 77843-3366, USA}

\begin{abstract}
Spatial correlation functions of quarkonia in the quark-gluon plasma are available from finite-temperature lattice-QCD with good precision. However, their interpretation in terms of microscopic spectral functions has been challenging due to a highly oscillating momentum integral in their Fourier transform and the need for a reliable evaluation of their 3-momentum dependence. 
Utilizing the thermodynamic $T$-matrix approach we first obtain a Lorentz-covariant scattering equation by taking advantage of a separable-potential approximation. Medium effects are included through state-of-the-art heavy-quark selfenergies and a potential screening which we constrain by euclidean correlators from lattice QCD. In addition to the usual two-particle cut in the scattering equation we also include particle-hole excitations, also referred to as zero mode.
We find that both the increase of euclidean correlators with 3-momentum and the suppression of the spatial correlators found in lattice-QCD can be qualitatively reproduced, with the zero mode playing an essential role for the former but not for the latter. 
\end{abstract}

\date{\today}
\maketitle

\section{Introduction}
\label{intro}
Heavy quarkonia ($Q\bar{Q}$) are key probes of how the fundamental force of Quantum Chromodynamics (QCD) is modified in the Quark-Gluon Plasma (QGP). On the one hand, the production systematics of quarkonia in ultra-relativistic heavy-ion collisions (URHICs) has been extensively studied in transport approaches to extract pertinent transport parameters, in particular the inelastic reaction rates that govern the kinetics of quarkonium dissociation and their regeneration from heavy quarks and antiquarks diffusing through the QGP. On the other hand, first-principles information on in-medium properties of quarkonia is obtained from lattice-QCD (lQCD) computations, which, however, is not directly deployable to transport calculations in heavy-ion collisions. In particular, real-time information such as spectral functions and the energy-momentum dependence of transport coefficients are not easily extracted from lQCD, typically requiring an inverse integral transform with a finite number of lattice ``data". Microscopic models and/or effective-field theory calculations have been utilized to bridge this gap~\cite{Rapp:2008tf,Mocsy:2013syh}. By making an ansatz for the interaction which generally involves a few parameters, one can calculate the Laplace transform from euclidean space into Minkowski space-time straightforwardly and thus constrain the model parameters. For quarkonia this strategy has been widely implemented using potential  models~\cite{Wong:2004zr,Alberico:2006vw,Rapp:2008tf,Mocsy:2013syh}. In particular, when deployed in momentum space, one arrives at a Lippmann-Schwinger equation for the two-body scattering amplitude~\cite{Cabrera:2006wh} that further allows for a seamless implementation into quantum many-body theory~\cite{kraeft1986quantum,Riek:2010fk,Liu:2017qah}, with self-consistently determined one- and two-body Green's functions as required for a strongly interacting system. In the past, lQCD constraints from heavy-quark (HQ) free energies and euclidean time correlators have been applied, and more recently from Wilson line correlators~\cite{Tang:2023tkm} and  extended operators~\cite{Tang:2024dkz} which enhance the sensitivity to the soft physics at the thermal scale.
There also exists lQCD results for spatial meson correlation functions which depend on the center-of-mass position of the bound-state (usually aligned in $z$ direction due to rotational invariance of the medium)~\cite{Ding:2012pt,Karsch:2012na,Bazavov:2014cta}. In principle, they carry a decisive advantage over euclidean correlators since the latter's $\tau$ variable is limited by the inverse temperature while the $z$-variable of the  spatial correlators is not and is therefore more sensitive to long-distance physics. However, since the conjugate variable is the total momentum, $P$, of the bound state, two main challenges for microscopic analyses arise. First, a reliable calculation of the $P$-dependence in potential models is not straightforward since the latter are not a priori Lorentz-covariant in the vacuum which induces spurious $P$-dependencies that may obscure medium effects. Second, the Fourier transform requires the evaluation of a highly oscillating integral for $P\to\infty$ with typically slow convergence. Recent progress on the latter issue has been made for light-flavor sector with a new technique that involves schematic ans\"atze for the meson spectral functions~\cite{Lowdon:2022xcl}. 
Here we present the first microscopic model calculation for the spatial correlators of quarkonia, thereby overcoming both aforementioned challenges. 

In the remainder of this paper we first set up a Lorentz-covariant quarkonium $T$-matrix in vacuum and then compute its medium modifications (Sec.~\ref{sec_tmat}). Next, we evaluate in-medium spectral functions (Sec.~\ref{sec_spec-func}) including the usual 2-body unitarity cut and the zero-mode contribution. This is followed by an analysis of euclidean correlators at finite 3-momentum (Sec.~\ref{sec_eucl-corr}) and spatial correlators (Sec.~\ref{sec_spatial}), including comparisons to lQCD data. We conclude in Sec.~\ref{sec_concl}.

\section{$T$-matrix at Finite 3-Momentum}
\label{sec_tmat}
We start from the covariant four-dimensional Bethe-Salpeter (BS) equation for the two-body scattering amplitude,
\beq
T(P;p',p) = V(p',p) + \int\frac{d^4k}{(2\pi)^4} V(p',k) \GQQb(P,k) T(P,k,p)
\label{BSE}
\eeq
with a Lorentz-covariant interaction potential $V$ and a heavy-quark-antiquark propagator, $\GQQb$. The total and relative four-momenta of the incoming and outgoing quarks are denoted by $P=(E,\vec P)$, $p=(p_0,\vec p)$ and $p'=(p_0',\vec p')$, respectively; we define the incoming anti-/quark 4-momenta as $p_\pm = P/2\pm p = (E/2\pm p_0, \vec P/2 \pm \vec p)$. 
For the purpose of the present paper, we will not employ a (Lorentz non-covariant) Cornell potential, but a simplified version based on a separable interaction that does not depend on the total 4-momentum, 
\bea
    V(p',p) &=& C v(p)v(p')
    \\
    v(p) &=& \left( \frac{2\Lambda^2}{2\Lambda^2 + 4|p^\mu p_\mu|} \right)^2 \ , 
\label{pot}
\eea
where $C$ denotes the coupling constant and $\Lambda$ is a momentum cutoff parameter to characterize the finite size of the interaction thereby regularizing the scattering integral.
A three-dimensional (3D) reduction of the BS equation that takes advantage of the suppressed energy transfer in the scattering of heavy particles but maintains the Lorentz-invariance of the $T$-matrix can be achieved in the Blankenbecler-Sugar reduction scheme~\cite{Blankenbecler:1965gx}, which, in particular, preserves the correct cut structure of the 2-body propagator~\cite{Brockmann:1996xy}. The relative energies are set to their on-shell values, \ie,  
$k_0=(\omega_{k_+}-\omega_{k_-})/2$, where $\omega_{k_\pm}^2 = m_Q^2 + (\vec P/2\pm \vec k)^2$, and after projecting on the anti-/particle states one obtains
\bea
\GQQb(E,\vec P;\vec k) = \frac{2m_Q^2(\omega_{k_+} + \omega_{k_-})  } {\omega_{k_+}\omega_{k_-}(s+\vec P^2 - (\omega_{k_+}+\omega_{k_-})^2 + i \epsilon)} 
\nonumber \\
\delta(k_0-(\omega_{k_+}-\omega_{k_-})/2)  \ , \hspace{2.5cm} 
\label{G2vac}
\eea
The same on-shell prescription is applied to the relative energies of the external quarks, $p_0=(\omega_{p_+} - \omega_{k_-})/2$.
The 3D scattering equation then takes the form
\begin{eqnarray}
T(P;\vec p',\vec p) &=& V(\vec p,\vec p') + \int \frac{d^3k}{(2\pi)^3} V(\vec{p}',\vec{k})
\nonumber\\
&& \hspace{1.5cm} \times G_{Q\bar{Q}}(P;\vec{k}) T(P;\vec{k},\vec{p}) \ .
\label{LS}
\end{eqnarray}
which can be resummed via a geometric series yielding
\beq
T(E,P;p,p') = \frac{V(\vec{p},\vec{p}')}{1-\int \frac{d^3k}{(2\pi)^3} G_{Q\bar{Q}}(E,P;k)V(\vec{k},\vec{k})} \ .
\label{Tmat-sol}
\eeq
We have numerically verified that this equation is independent of the total three-momentum, $\vP$, \ie, depending only on $s=E^2-\vP^2$. 
This further implies that the separable potential, eq.~(\ref{pot}), carries no further angular dependence and thus describes only $S$-wave channels. We will neglect any spin splittings and focus on the charmonium case, 
with a ground-state mass of $M_{\eta_c} = 2.98 \ \text{GeV}$
and  a schematic vacuum width in line with the empirical value for $\eta_c$ meson, $\Gamma_{\eta_c}\simeq$30\,MeV; the latter does not figure in our in-medium calculations reported below.

\subsection{In-Medium $T$-Matrix}
\label{ssec_tmat-med}
When embedding quarkonia into QCD matter, two principal medium effects occur, \ie, a dressing of the anti-/quarks in the two-body propagator amounting to selfenergies due to interactions with the thermal partons from the QGP, and Debye screening of the potential (as well as interference effects giving rise to an $r$-dependent width which in the long-distance limit correspond to twice the individual HQ widths).
In principle, both effects can be computed self-consistency as has been done in previous applications of the $T$-matrix formalism with an in-medium Cornell potential as input~\cite{Liu:2017qah,Tang:2023tkm}. 
With the schematic input potential used here, such a calculation is not warranted and we therefore implement the two medium effects independently but guided by earlier results. 

In the sQGP, with its inherently large off-shell effects manifest in large collisional widths and associated broad spectral functions, we compute the energy loop integral of the two-body propagator from the underlying HQ spectral functions. Using the imaginary-time formalism of finite-temperature field-theory one obtains two contributions. The first one is the two-body ``unitarity cut" whose imaginary part can be written as
\begin{eqnarray}
\text{Im}G_{Q\bar{Q}}(P) &=& -\frac{1}{2}\int\limits_{-E/2}^{E/2}\frac{d\omega}{2\pi} 
\rho_Q(\omega_+,\vec{k}_+)\rho_{\bar Q}(\omega_-,\vec{k}_-) 
\nonumber \\
 & & \qquad \times \left(1-f_Q(\omega_+)-f_{\bar Q} (\omega_-)\right)
\ .
\label{ImG2}
\end{eqnarray}
with $\omega_\pm = E/2\pm\omega$ and Fermi distribution functions, $f_Q$. 
The spectral functions $\rho_{Q,\bar Q}=-2{\rm Im}G_{Q, \bar Q}$   (which are identical at vanishing quark chemical potential $\mu_q=0$) are defined 
via the pertinent single-parton propagator,
\beq
G_Q(\omega,k) = 2m_Q /(\omega^2-k^2-m_Q^2-2m_Q\Sigma_Q(\omega,k;T)) \ ,  
\eeq
which in turn depend on the complex selfenergy, $\Sigma(\omega,k)$.
Selfconsistent thermodynamic $T$-matrix calculations predict that the HQ selfenergies develop large imaginary parts corresponding to collisional widths of around 0.5\,GeV~\cite{Liu:2017qah,Tang:2023tkm}, which are, in fact, required to generate HQ transport coefficients that are compatible with open HF phenomenology in URHICs~\cite{Krishna:2025bll}. These widths create very broad spectral functions which mandate the inclusion of pertinent off-shell effects in the 2-body propagator. Its real part is readily obtained from a dispersion relation, 
\begin{eqnarray}
    \text{Re}[G_{Q\bar{Q}}(s;k)] = \frac{1}{\pi}\mathcal{P} \int \frac{\text{Im}[G_{Q\bar{Q}}(s';k)]}{s'-s} ds' \ .
    \label{dispersion}
\end{eqnarray}
Here, we will employ the selfenergies from selfconsistent
$T$-matrix results based on constraints from HQ free energies and quarkonium correlators from lattice QCD within the ``strongly coupled scenario" (SCS)~\cite{Liu:2017qah}, but we also explore variations thereof as detailed below. 

The second contribution is a particle-hole excitation where an incoming heavy quark scatters off the medium (diagrammatically, this corresponds to turning around the heavy antiquark line of the quarkonium propagator). Its imaginary part takes the form~\cite{Riek:2010py}
\bea
{\rm Im} G_{ph}(P,\vec k) &=& -\int\limits_0^\infty 
 \frac{d\omega}{2\pi} \rho_Q(\omega,\vec k) \rho_Q(E+\omega, \vP+\vec k) 
 \nonumber  \\ 
 & & \quad \times [f_Q(\omega) - f_Q(E+\omega)]
\label{Gph}
 \eea
This contribution does not involve a $Q\bar Q$ $T$-matrix, but is generated by heavy-light scattering encoded in the HQ selfenergies. It is also known as the ``zero-mode" contribution, and its timelike limit is directly related to the HQ diffusion coefficient~\cite{Aarts:2005hg,Riek:2010py}.

The medium modifications of the separable potential are introduced via a Debye-like screening of the vertex function with strength parameter $a$, 
\begin{eqnarray}
    v(\vec{q};T) = \left( \frac{2\Lambda^2}{2\Lambda^2 + 4|q^\mu q_\mu| + (aT)^2} \right)^2 , 
    \label{in-med-pot}
\end{eqnarray}
which simulates a reduction of the interaction range. We also account for a temperature-dependent coupling constant, $C(T)$, which simulates an in-medium reduction of the interaction strength.
Our strategy in this exploratory calculation is to use the 3-momentum dependent euclidean correlators to constrain the in-medium interaction and then predict the spatial correlation functions from it.

\begin{figure}[!htb]
    \centering
    \includegraphics[width=0.99\linewidth]{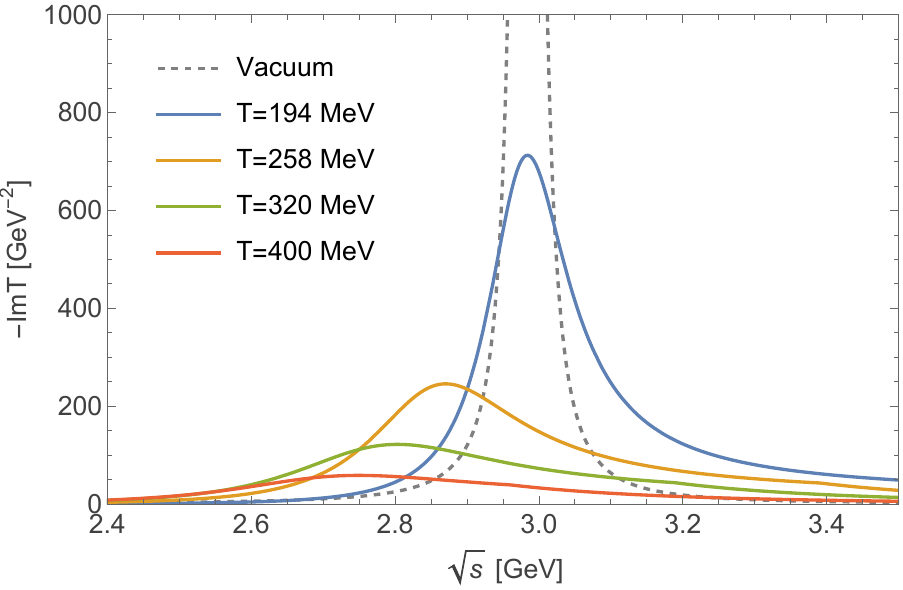}
    \caption{Imaginary part of the $S$-wave charmonium $T$-matrix at vanishing total 3-momentum in vacuum (gray dashed) and in the QGP at temperatures $T$=194, 258, 320 and 400 MeV (blue, orange, green and red lines, respectively).}
    \label{fig:tmat}
\end{figure}

\begin{figure*}[!htb]
    \centering
    \includegraphics[width=0.99\linewidth]{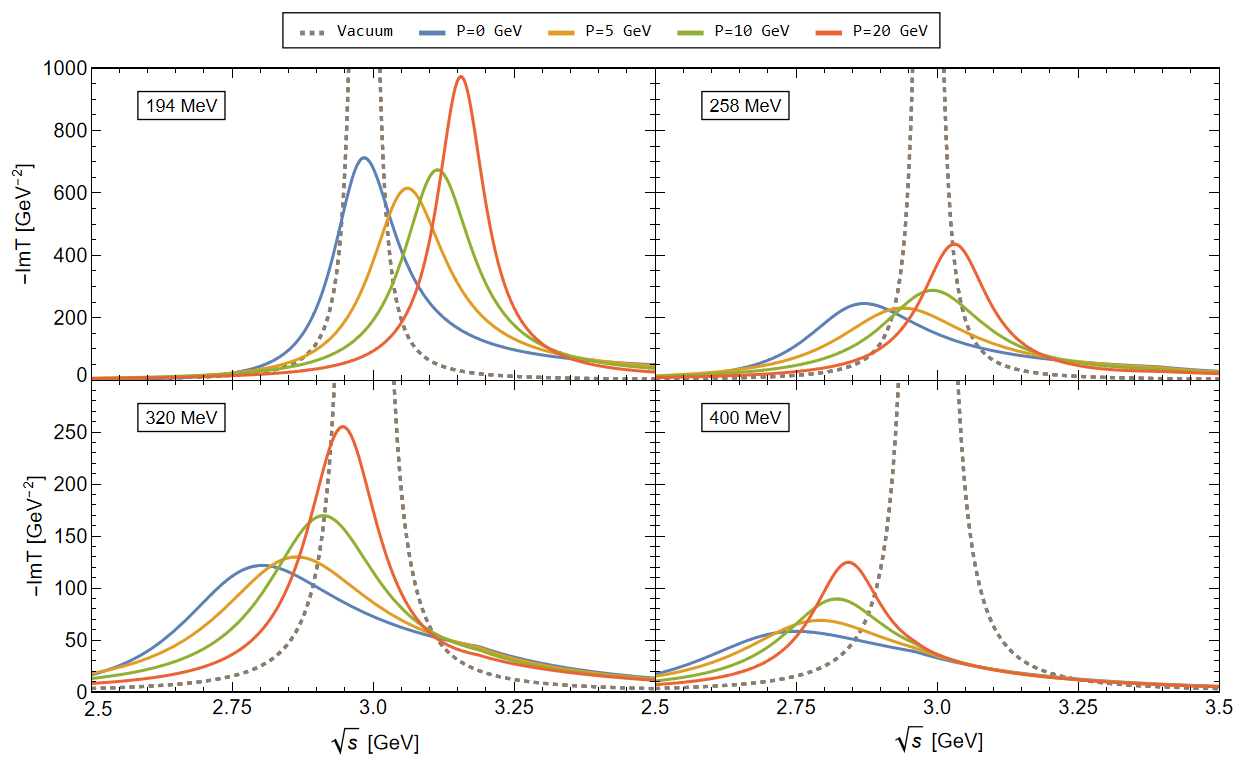}
    \caption{Imaginary part of the on-shell $S$-wave charmonium $T$-matrix for finite 3-momenta ($P$=0, 5, 10, 20\,GeV for blue, orange, green and red lines, respectively) at different temperatures $T$=194, 258, 320 and 400 from the upper left to lower right panel. Note the different $y$-scale in the upper vs. lower panels.} 
    \label{fig:tmat-P}
\end{figure*}

The resulting in-medium $T$-matrices for a typical parameter set of $\Lambda=2.4$\,GeV, $C = -54.94$\,GeV (which reproduces the vacuum ground-state mass) and  $a=1.7$ (whose choice will be detailed below) are shown in Fig.~\ref{fig:tmat}. One finds the expected strong in-medium broadening and a slight shift to lower masses at $P$=0. At finite momentum, compiled in Fig.~\ref{fig:tmat-P}, we find that the bound-state peaks shifts toward higher invariant masses indicating a loss in binding energy as the screening in the form factor is enhanced.
This is accompanied by a moderate initial broadening at low momenta at the lowest two temperatures followed by a narrowing at $P\gtrsim10$\,GeV, while for the 2 higher temperatures the narrowing occurs already at low momenta.

\section{Spectral Functions}
\label{sec_spec-func}
The spectral functions, $\sigma$= $-\frac{1}{\pi}$Im $\cG$, needed to evaluate correlators in euclidean space-time follow from two-point correlation functions, $\cG$, by closing the $T$-matrix with $Q\bar Q$ loops according to~\cite{Cabrera:2006wh}
\beq 
\cG = \cG_0 + \cG_0 T {\cG}_0 \ .
\eeq 
The non-interacting loop function is given by
\beq
    \cG_0(s,P)= N_fN_c\int\frac{d^3k}{(2\pi)^3} G_{Q\bar{Q}}(s,P;\vec{k})  \Tr[\Lambda_+\Gamma_M \Lambda_-\Gamma_M]
    \label{G0}
\eeq
where $\Gamma_M$ denotes the vertex in meson channel, $M$, and $N_c$=3, $N_f$=1. 
To leading order in $1/m_Q$, the trace in the pseudoscalar $\eta_Q$ channel 
($\Gamma_M=\gamma_5$) gives -$[(\omega_++\omega_-)^2 - P^2]/2m_Q^2$. In vacuum this corresponds to the $Q\bar Q$ continuum above the 2-particle threshold. For the rescattering term one has~\cite{Cabrera:2006wh}
\begin{eqnarray}
\begin{split}
    \Delta \cG(s,P) = N_f N_c\int \frac{d^3k}{(2\pi)^3}\frac{d^3k'}{(2\pi)^3} G_{Q\bar{Q}}(s,P;\vec{k}) 
    \\
    \times T(s,P;\vec{k},\vec{k}')G_{Q\bar{Q}}(s,P;\vec{k}') 
    \\
    \times \text{Tr}[\Lambda_+(\vec{k}_+)\Gamma_M\Lambda_-(\vec{k}_-)\tilde{\Gamma}\Lambda_-(\vec{k}_-')\Gamma_M\Lambda_+(\vec{k}_+')]
\end{split}
\end{eqnarray}
where we have also employed the leading order in the 1/$m_Q$-expansion yielding a value of 2 for the trace~\cite{Cabrera:2006wh}. Note that due to the separable nature of the potential, the $d^3k$ and $d^3k'$ integrals are identical and factorize. For the HQ selfenergy we implement numerically tabulated results from Ref.~\cite{Liu:2017qah} for the 4 temperatures available from the SCS~\cite{Liu:2017qah} (which results in large widths characteristic of a strongly coupled QGP). In particular, for the interacting part, which is dominated by the bound-state contribution, we allow for a suppression of 50\%  of the selfenergy to simulate the presence of interference effects (also known as imaginary part of the $\QQb$ potential), which technically correspond to the presence of 3-body correlations between the medium parton and the $Q$ and $\bar Q$. The choice of the 50\% coefficients results in spectral functions  whose bound-state peak height is quantitatively quite similar to the original calculations when the interference effect is accounted for, cf.~Fig.~9 in Ref.~\cite{Liu:2017qah}. For the (uncorrelated) continuum as well as the zero-mode contribution these effects are absent and the full selfenergy is used.

\begin{figure}[!th]
    \centering
    \includegraphics[width=0.99\linewidth]{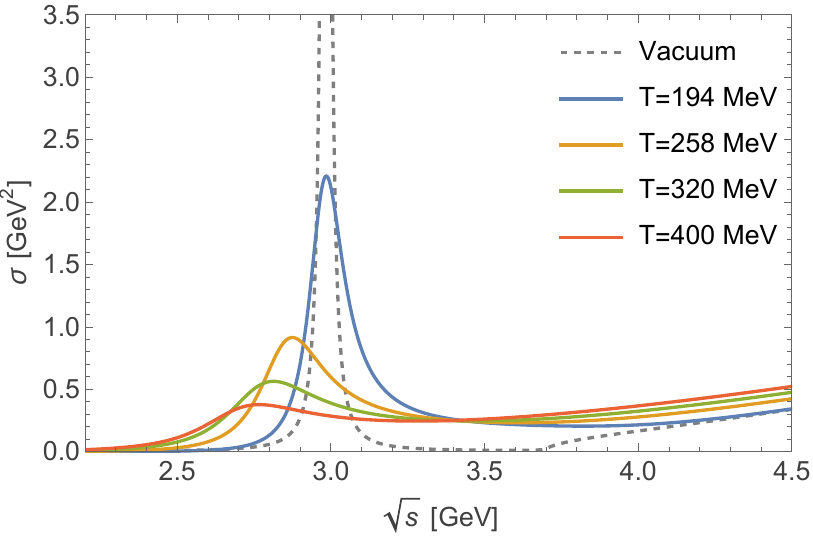}
    \caption{Temperature dependence of $S$-wave charmonium spectral functions at vanishing total 3-momentum obtained from our in-medium $T$-matrix with separable potential as shown in Fig.~\ref{fig:tmat}.}
    \label{fig:spec}
\end{figure}

\begin{figure*}[!th]
    \centering
    \includegraphics[width=0.99\linewidth]{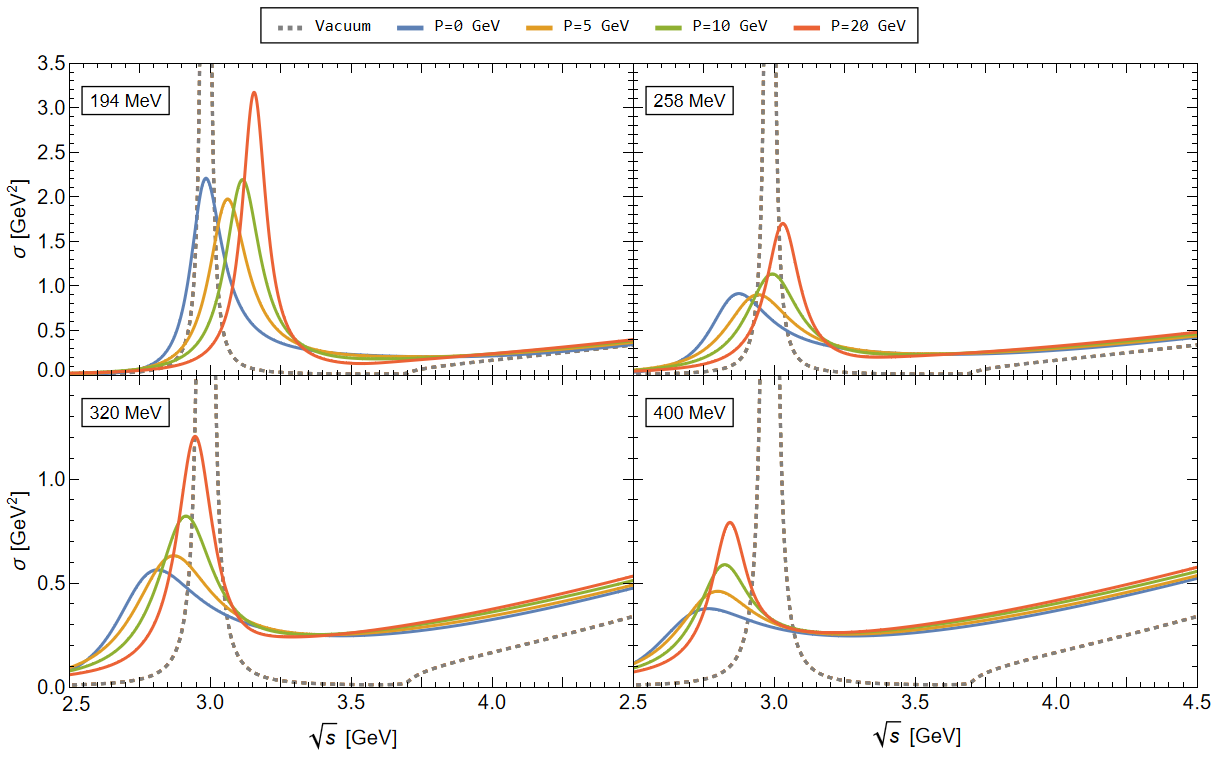}
    \caption{Three-momentum dependence of $S$-wave charmonium spectral functions at different temperatures $T$=194, 258, 320 and 400\,MeV (from top left to bottom right panel), obtained from the in-medium $T$-matrices shown in Fig.~\ref{fig:tmat-P} with the same color identification for the different momenta.}
    \label{fig:spec-P}
\end{figure*}

The resulting spectral functions, shown in Figs.~\ref{fig:spec} and \ref{fig:spec-P}, closely mirror the imaginary part of the $T$-matrices in the bound-state region, but additionally feature an in-medium continuum that in the near threshold-region is markedly enhanced over its vacuum form due to the broad HQ spectral functions.
%
%

\begin{figure}[!th]
    \centering
    \includegraphics[width=0.99\linewidth]{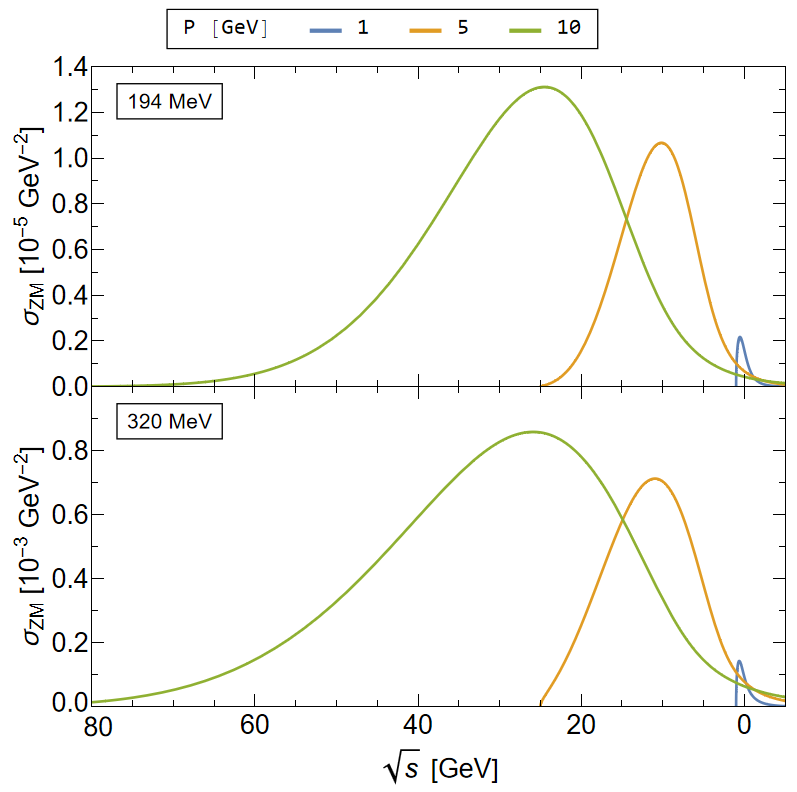}
    \caption{Zero-mode contribution to the charmonium spectral functions at finite momentum for $T$=194 and 258\,MeV.}
    \label{fig:zm}
\end{figure}

The zero-mode (ZM) contribution to the spectral function is given by
\beq
\sigma_{ZM}=-\frac{N_c}{\pi}\int \frac{d^3k}{(2\pi)^3}\text{Im}G_{ph}(P,\vec{k})\Tr[\Lambda_+\Gamma_M\Lambda_+\Gamma_M^\dagger] \ .
\eeq
In the $\eta_c$ channel the trace evaluates to   
\begin{eqnarray}
    \Tr[\cdots] =\frac{\omega_k\omega_{k+P} -(|\vec{k}|^2 + \vec{k}\cdot\vec{P}+m_Q^2)}{m_Q^2} \ ,
\end{eqnarray}
which vanishes for $\vP$=0 but not for finite $\vP$~\cite{Aarts:2005hg,Riek:2010py}, cf.~Fig.~\ref{fig:zm}. Also note that it is mostly concentrated in the space-like region (albeit at positive energy), and while its magnitude is much smaller than that for the unitarity cut, its contribution to the correlators will be magnified by a thermal factor in the Laplace transform.

\section{Euclidean Correlators}
\label{sec_eucl-corr}
The Euclidean-time correlator is related to the spectral function through a Laplace transform,  
\beq
G(\tau; T,P) =\int_{0}^\infty \frac{dE}{2\pi} \ K(\tau,E,P;T) \sigma(E,P;T)
    \label{ec} \\
\eeq
with the heat kernel
\beq
K(\tau,E,P;T)  =\frac{\cosh(E(\tau - \beta/2))}{\sinh(E\beta/2)} \label{kernel}
\eeq
As is commonly done in lQCD calculations~\cite{Ding:2012pt}, we consider euclidean correlator ratios (ECRs):
\begin{eqnarray}
    R_G(\tau;T) = \frac{G(\tau; T,P)}{G_{\text{rec}}(\tau;T,P)}
\end{eqnarray}
where the denominator is a ``reconstructed" correlator with a low-temperature spectral function as the reference,
\begin{eqnarray}
    G_{\text{rec}}(\tau;T,P) = \int_{0}^\infty \frac{dE}{2\pi} \ K(\tau,E,P;T) \sigma(E,P;T_{\rm ref}) \ .
\end{eqnarray}
In our calculations, we use the spectral functions at the lowest available temperature of $T$=194\,MeV from Ref.~\cite{Liu:2017qah} for the reconstructed correlator.


\begin{figure}[!h]
    \centering
    \includegraphics[width=0.95\linewidth]{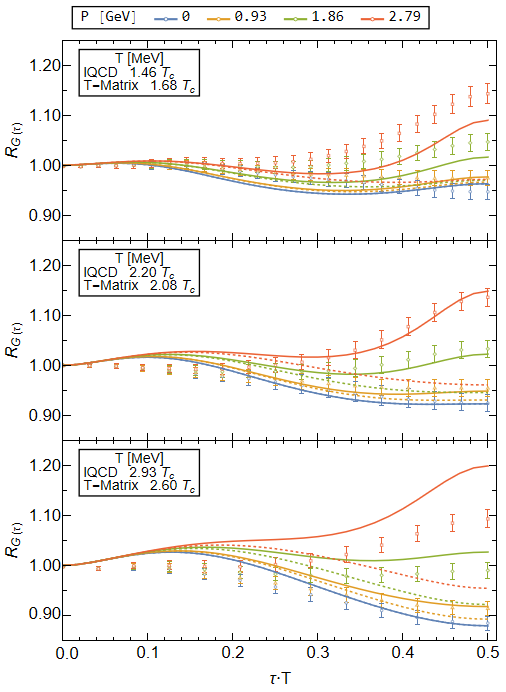}
    \caption{$T$-matrix results for ECRs for $T$=258, 320 and 400\,MeV  (top to bottom), compared to quenched lQCD data~\cite{Ding:2012pt} at finite momenta with dashed lines representing the unitarity cut and solid lines the full result including the zero mode.}
    \label{fig:ecrs}
\end{figure}

As mentioned above, our strategy is to utilize the ECRs at $P$=0 to constrain the medium modifications of the potential parameters. We carry that out iteratively by first selecting a value for the momentum cut-off $\Lambda$ which then determines the vacuum coupling constant $C$ to reproduce the vacuum ground-state mass. We then select a Debye-mass coefficient, $a$, and take the vacuum coupling as an upper bound for the lowest temperature, which is also chosen to reproduce the same bound-state mass (as found in Ref.~\cite{Liu:2017qah}). We then successively adjust the coupling at the next highest temperatures (treating the previous as an upper bound) to quantitatively reproduce the large-$\tau$ limit of the lQCD results in Ref.~\cite{Ding:2012pt}. This is repeated until a set of parameters is found which minimizes strong deviations from the lQCD data.
With this set-up we then move on to obtain the 3-momentum dependent ECRs, including the ZM contribution, as predictions. 

Our results are summarized for 3 temperatures in Fig.~\ref{fig:ecrs} and compared to the lQCD results of Ref.~\cite{Ding:2012pt}.
While there are also unquenched lQCD results available at $P$=0 (which are qualitatively similar, but with a smaller suppression below one~\cite{Aarts:2007pk}), we chose the former, since they are also available at finite $P$. We also do not have an accurate temperature matching between the quenched lQCD correlators and our selfenergy results. However, using the $P$=0 result as the benchmark for our in-medium effects, the dependence on 3-momentum at each temperature and its progression with temperature should give us some qualitative insights. Indeed, the main feature of the finite-$\vP$ ECRs from lQCD of a significant enhancement of the large $\tau$ growing with momentum. Our calculations qualitatively (even semi-quantitatively) reproduce this enhancement which turns out to be mainly due to the ZM contribution. At higher temperatures, the large-$\tau$ enhancement seen in the the lattice results becomes increasingly over-estimated by our calculation, with a significant portion originating from .
the unitarity cut.


\section{Spatial Correlators}
\label{sec_spatial}
The spatial correlator can be expressed through the spectral function as 
\beq
    \tilde{G}(z;T) = \int_0^\infty \frac{dE}{\pi E} \int \frac{dP}{2 \pi} \ \text{Exp}(iPz) \ \sigma(E,P;T) \ .
\label{sc}
\eeq 
For large $z$ the integrand is a rapidly oscillating function in $P$ whose frequency is given by $z$ which poses a formidable numerical challenge. 

By using schematic spectral functions as a test case, we have determined that for the momentum integration to converge it must be carried out to order $P \simeq1$\,TeV, far exceeding the physics reach of our model and becoming numerically unfeasible. This is a direct consequence of the oscillating kernel, which drives the convergence through large cancellations across individual periods. However, for sufficiently large momentum, the spectral function recovers Lorentz invariance. However, we have found that selecting a momentum cutoff value of $P_{\rm cut} = 250$\,GeV above which we approximate the spectral function to be momentum independent, can solve this problem, since including momentum dependence above this value has a negligible  effect on the correlator. In particular, we have verified that quantitatively similar results as shown here are obtained with smaller cutoff values. This stands in contrast to the numerical requirements of the ECRs, where an intergration range up to the order of $\sqrt{s} \simeq 100 \ \text{GeV}$ as $\tau \rightarrow 0$ is sufficient. Indeed, for large $\sqrt{s}$ values the spectral function recovers the vacuum result. In practice, we select a cutoff value of $\sqrt{s} = 23.5 \ \text{GeV}$ above which we set the spectral function equal to the non-interacting vacuum spectral function, $\sigma = \frac{3s}{8\pi^2}\sqrt{1-\frac{4m_c^2}{s}}$. We have verified that extending the numerical calculation out to higher $\sqrt{s}$ has little effect on the ECRs in all $\tau$ regions.

\begin{figure}[H]
    \centering
    \includegraphics[width=0.95\linewidth]{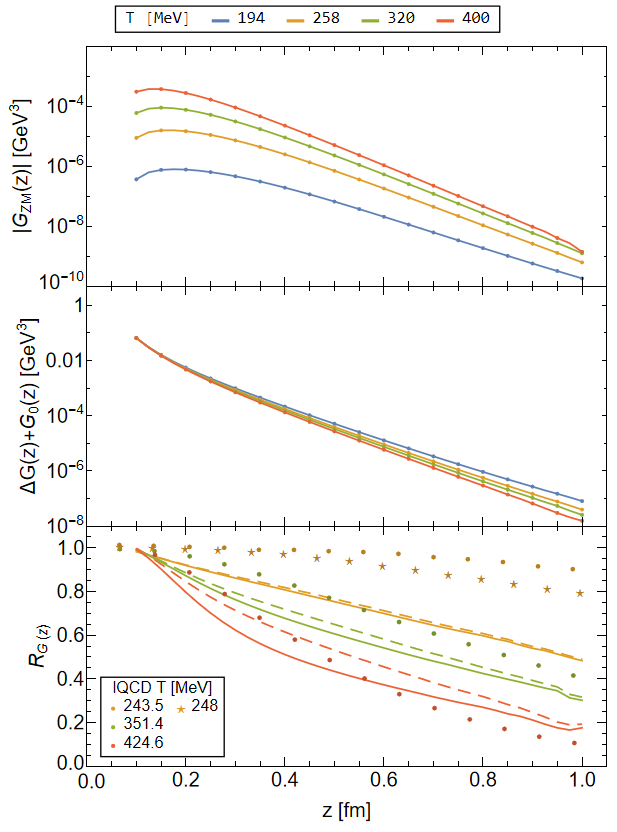}
    \caption{Spatial correlators of $S$-wave charmonium in the QGP from the zero-mode (absolute magnitude, upper panel) and the unitarity cut (including both free and interacting parts, middle panels) and their ratio to the reconstructed correlator compared to the lQCD data (dots) of Ref.~\cite{Karsch:2012na} (bottom panel). In the lower panel, dashed lines represent the unitarity cut contribution only, while the solid lines include also the ZM. }
    \label{fig:scrs}
\end{figure}

For large $z$, we expect the correlator to take the form $G(z) \approx A_1\text{Exp}(-m_1 z) + A_2 \text{Exp}(-m_2z) + ...$, where $m_1 < m_2< ...$, and each $m_i$ corresponds to the screening mass of a specific excitation channel~\cite{Lowdon:2022xcl}. In the calculation, we only consider the $\eta_c$ channel featuring a single resonance state, and thus $G(z) \approx A_{\eta_c}\text{Exp}(-m_{\eta_c}z)$. Indeed, as shown in Fig.~\ref{fig:scrs}, at large $z$, each correlator has the expected exponential shape, implying that the numerical calculation was carried out with sufficient stability and momentum range.

Turning to the SCRs, we calculate the ratio $R_G(z)$ with a reconstructed correlator in the denominator using our lowest-temperature reference. Our results are shown in Fig.~\ref{fig:scrs}. 
%
%
 %
%

In the upper panels 2 panels we show the temperature dependence of the calculated spatial correlation functions for the zero-mode (top) and the unitarity cut (middle). While the latter has a rather mild temperature dependence with a hierarchy that only develops at rather large $z$ and consists of an increasing suppression with temperature, the zero mode produces a much stronger and opposite temperature dependence, especially at low $z$ (note, however, that the the sign of the zero-mode correlator is negative; this can also be shown analytically in the zero-width limit for the HQ selfenergies). The absolute magnitude of the zero-mode contribution is subleading at any $z$ compared to the unitarity cut, as can be seen more clearly from the dashed vs.~the solid lines in the correlators ratios shown in the lower panel on a linear scale. This is similar (albeit it with opposite sign) to the up to ca.~20\% enhancement found in the ECRs at large $\tau$, which suggests that the intermediate-$z$ region around $\sim$0.5\,fm probes similar physics as the large $\tau$ regime of the ECRs (which corresponds to comparable values of $\tau$).
We also compare to the lattice data for $\eta_c$ correlators in $N_f$=2+1-flavor QCD from Ref.~\cite{Karsch:2012na} where we indicate the absolute temperatures for the 3 datasets which turn out to be within 10\% of the temperatures used in our calculation.  At the lowest temperature we overpredict the suppression found in lQCD, but the agreement somewhat improves at higher $T$. 
It may be noteworthy that the in-medium potential extracted for the SCS, which is at the basis of our HQ selfenergies, exhibits a rather appreciable screening with temperature. On the other hand, more recent constraints based on Wilson Line correlators result in less potential screening~\cite{Tang:2023tkm} which might reduce the suppression in the SCRs at intermediate temperatures.
In a more recent lQCD evaluation for $N_f$=2+1 with near realistic pion mass ($m_\pi$=160\,MeV)~\cite{Bazavov:2014cta} $\eta_c$ SCRs  
have been computed for temperatures up to 248 MeV. The suppression at high $z$, down to about 0.8, is somewhat larger than in the lQCD data shown in the figure, but still not as large as in our calculation. On the other hand, the lQCD SCRs  for $T$=197\,MeV in the more recent computation are very close to 1, supporting the use of our $T$=194~MeV results as the reconstructed correlator. 

\section{Conclusions}
We have employed the thermodynamic T-matrix formalism to compute the 3-momentum dependence of $S$-wave quarkonium spectral functions in the QGP. To construct a manifestly Lorentz-invariant vacuum 
interaction -- required to avoid any spurious momentum dependence in medium -- we started from the 4D Bethe Salpeter equation in connection with a rank-one separable potential and fitted its 2 parameters to the ground-state charmonium mass.
We then deployed this setup to the QGP adopting heavy-quark selfenergies extracted from previous lQCD-constrained selfconsistent results in a strongly coupled QGP~\cite{Liu:2017qah}. 
The medium effects on the potential were introduced through a Debye screening in the formfactor and a temperature-dependent coupling constant and constrained by euclidean-time correlators at zero total momentum from thermal lattice QCD. The results at finite $P$ can then be considered as a prediction our the approach. With increasing $P$, the pertinent $T$ matrices feature, for the most part, a narrowing of the bound-state peak along with an increase of its pole position.  

The corresponding spectral functions closely mirror the behavior of the $T$-matrices, but, in addition to the standard two-particle ($\QQb$) cut (unitarity cut), pick up a particle-hole (or zero-mode) part dictated by thermal field theory. The zero-mode is generated by $Q\to Q$ scattering off the QGP, which is encoded in the HQ selfenergy through heavy-light scattering amplitudes and does not involve any extra parameter. It turns out that the zero-mode plays a central role in enhancing the large-$\tau$ range of the finite-momentum ECRs, at a level of 10-20\%, which turns out to be semi-quantitatively in line with (quenched) lQCD data. 
The evaluation of spatial correlators required substantial care in their numerical variation due to a highly oscillating integrand. We succeeded to control the numerics out to distances of about $z$=1\,fm, and found a substantial suppression with increasing temperature of their ratios to the lowest-temperature correlator. Again, qualitative agreement with (unquenched) lQCD computations has emerged. The zero-mode still affect the SCRs at the 10-20\% level, which, however, is on top of a much larger effect from the in-medium bound-state and thus less prominent. Understanding the precise origin of the SCR suppression remains a challenge, but at this point the rather uniform upward mass shift of the bound state with $P$ across different temperatures appears to be a reasonable candidate for this behavior. Further developments of implementing a more realistic interaction (\eg, a higher-rank separable potential or a suitable modified Lorentz-covariant Cornell potential)and recently updated HQ selfenergies~\cite{Tang:2023tkm} are warranted to gain deeper insights into this question. In addition, applications to the light-flavor sector seem promising, especially in light of recent work~\cite{Lowdon:2022xcl} which inferred  the persistence of color-singlet ``thermo-particles" deep into the QGP.

\label{sec_concl}

\acknowledgments
We thank Peter Lowdon, Owe Philipsen and Peter Petreczky for valuable discussion, and Isaac Sarver for his contributions early on in this project. This work is supported by the U.S. National Science Foundation under grant nos.\,PHY-2209335 and PHY-2514775, and by the Department of Energy via the Topical Collaboration in Nuclear Theory on \textit{Heavy-Flavor Theory (HEFTY) for QCD Matter} under award no.\,DE-SC0023547.

\bibliographystyle{unsrt}
\bibliography{references}

\end{document}